# Multi-spin-assisted optical pumping of bulk $^{13}$C nuclear spin polarization in diamond

Daniela Pagliero[1], K. R. Koteswara Rao[1], Pablo R. Zangara[1], Siddharth Dhomkar[1],
Henry H. Wong[1], Andrea Abril[3], Nabeel Aslam[4], Anna Parker[6], Jonathan King[6], Claudia E. Avalos[6],
Ashok Ajoy[6], Joerg Wrachtrup[4,5], Alexander Pines[6,7], and Carlos A. Meriles[1,2,†]

[1]Department of Physics, CUNY-City College of New York, New York, NY 10031, USA
[2]CUNY-Graduate Center, New York, NY 10016, USA
[3]Department of Physics, Universidad Nacional de Colombia, Bogotá D.C., Colombia
[4]3rd Physics Institute, University of Stuttgart, 70569 Stuttgart, Germany
[5]Max Planck Institute for Solid State Research, 70174 Stuttgart, Germany
[6]Department of Chemistry, University of California, Berkeley, California 94720, USA
[7]Materials Sciences Division, Lawrence Berkeley National Laboratory, Berkeley, California 94720, USA

One of the most remarkable properties of the nitrogen-vacancy (NV) center in diamond is that optical illumination initializes its electronic spin almost completely, a feature that can be exploited to polarize other spin species in their proximity. Here we use field-cycled nuclear magnetic resonance (NMR) to investigate the mechanisms of spin polarization transfer from NVs to $^{13}$C spins in diamond at room temperature. We focus on the dynamics near 51 mT, where a fortuitous combination of energy matching conditions between electron and nuclear spin levels gives rise to alternative polarization transfer channels. By monitoring the $^{13}$C spin polarization as a function of the applied magnetic field, we show $^{13}$C spin pumping takes place via a multi-spin cross relaxation process involving the NV$^-$ spin and the electronic and nuclear spins of neighboring P1 centers. Further, we find that this mechanism is insensitive to the crystal orientation relative to the magnetic field, although the absolute level of $^{13}$C polarization — reaching up to ~3% under optimal conditions — can vary substantially depending on the interplay between optical pumping efficiency, photo-generated carriers, and laser-induced heating.

## I. INTRODUCTION

The practice of nuclear magnetic resonance (NMR) presently encompasses multiple disciplines spanning analytical and medical sciences, biochemistry, environmental monitoring, and well logging, to mention just a few. A limitation common to all these applications, however, is the sample spin polarization, typically a minute fraction of the possible maximum. Sensitivity restrictions set a limit on the minimum amount of sample that can be detected (a concern when analyzing mass-limited systems or rare molecular moieties in solution) and result in longer acquisition times. Resorting to cryogenic temperatures or stronger magnets are the most obvious paths to enhanced spin polarization (and hence improved detection sensitivity), but sample freezing is often impractical (consider, e.g., living organisms) and large magnets tend to be disproportionately expensive.

Adding to the existing library of free-radical-based dynamic nuclear polarization (DNP) schemes[1], the use of optically polarized paramagnetic defects in wide bandgap semiconductors is emerging as an alternative polarization enhancement route of growing interest. An example of prominent importance is the negatively-charged nitrogen-vacancy (NV$^-$) center in diamond, a spin-1 system formed by a substitutional nitrogen and an adjacent vacancy. Green illumination efficiently pumps NV$^-$ into the $m_S = 0$ state of the ground triplet via spin-selective intersystem crossing[2]. This feature has already been exploited to demonstrate record levels of room temperature $^{13}$C spin polarization at low[3-5], intermediate[6], and high magnetic fields[7]. Recent studies based on other defects in diamond[8] or wide-bandgap host crystals other than diamond[9] have led to comparable results.

Here we focus on the dynamics of optically pumped $^{13}$C spin polarization in diamond near 51 mT, a regime with alternative polarization channels where, nonetheless, a clear understanding of the mechanics at play is still missing. Through the combination of field-cycled NMR, optically detected magnetic resonance (ODMR), and numerical modeling we unambiguously show that $^{13}$C nuclei polarize through multi-spin processes involving the electronic spin of the NV$^-$ and the electronic and nuclear spins of substitutional nitrogen (also known as P1 centers). We attain positive or negative nuclear spin polarization of up to 3% under ambient conditions, approximately a 3000-fold enhancement over the thermal $^{13}$C polarization at 9.4 T. We observe similar levels of $^{13}$C polarization over a broad range of crystal orientations if the magnetic field amplitude is changed so as to properly match the NV$^-$ and P1 center electron spin transitions. On the other hand, comparing the NMR signal amplitudes under different illumination conditions, we conclude the $^{13}$C polarization varies significantly with the beam diameter, an effect we associate to photo-induced charged processes and temperature changes in the diamond crystal. During the writing of this manuscript we learned that similar experiments have been carried out recently by Meijer and collaborators[10].

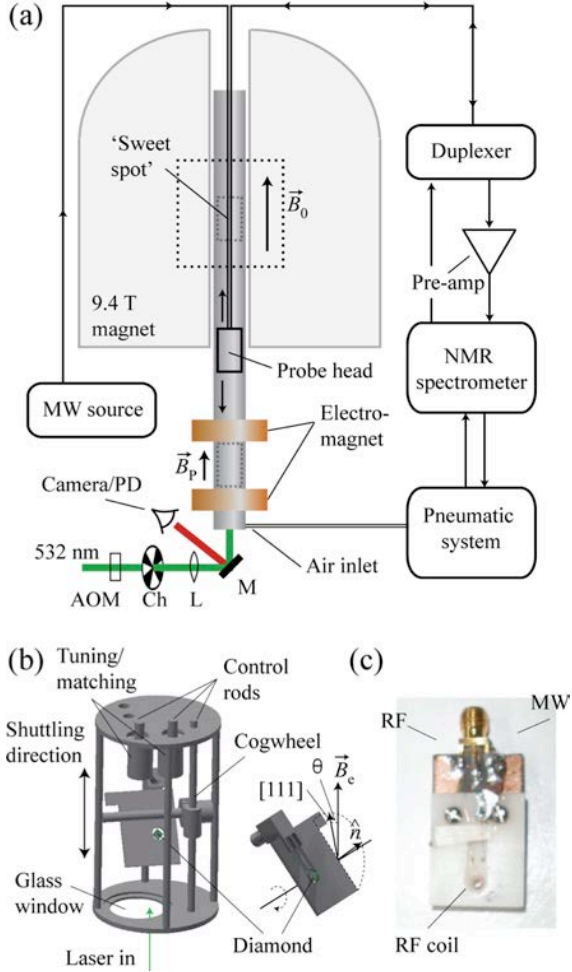

**Fig. 1:** (a) Schematics of the experimental setup. We use a pneumatic mechanism to shuttle the NMR probe head between the polarization field $B_P$ and the 9.4 T detection field $B_0$. Green laser light is used to pump $^{13}$C nuclear spin polarization at $B_P$. (b) Schematics of the probe head. A cogwheel system and a control rod (not shown) allow us to vary the angle $\theta$ between the [111] crystal axis and $B_P$. In the drawing $\hat{n}$ denotes the normal to the diamond surface. (c) Photograph of the diamond crystal, RF excitation/pick-up coil, and sample holder. A MW look in the back (not visible) allows us to resonantly excite the NV and P1 centers. AOM: acousto-optic modulator; L: lens; M: mirror; Ch: chopper; PD: photo-detector; RF: radio-frequency; MW: microwave.

## II. EXPERIMENTAL RESULTS

The sample we use is a Type 1b, [100] diamond crystal (3.2×3.2×0.3 mm$^3$) with a nitrogen content of 70 ppm. The preparation protocol comprises high-energy electron irradiation (7 MeV at a dose of $10^{18}$ cm$^{-2}$) and subsequent thermal annealing (2 hours at 700 $^0$C) resulting in an NV$^-$ concentration of about 10 ppm. Fig. 1 shows a schematic of our experimental setup: The diamond sample sits on a larger (1.5×1.5 cm$^2$) sapphire crystal, itself integrated into a sample holder within a customized NMR probe head featuring a microwave (MW) loop and a 3 mm diameter radio-frequency (RF) coil. The latter is part of an excitation/detection tank circuit completed by a pair of tuning/matching variable capacitors. We use a custom-made pneumatic system to shuttle the probe head between the 'sweet spot' of a 9.4 T superconducting magnet and a terminal point outside the bore where the magnetic field $B_P$ can be adjusted to values near 51 mT with the help of a split electromagnet. The probe head also features a cogwheel system and an external control rod allowing us to change the crystal orientation relative to the external field (pointing along the magnet bore axis). We use a set of lenses to focus the laser beam on the sample surface, and rely on an acousto-optic-modulator (AOM) to control the illumination on/off times. We can directly image the sample fluorescence with the aid of a camera and a suitable long-pass optical filter; when necessary, we use a mechanical chopper (Ch), a photo-detector (PD), and a lock-in amplifier (not shown) to measure the NV$^-$ fluorescence under optical excitation.

In a typical run, dynamic polarization of the $^{13}$C spins takes place for a variable time $t_L$ at a 'pumping' field $B_P$, after which the probe head is shuttled to the inside of the high-field magnet for detection (Fig. 2a); the shuttling time is approximately 1.5 s. An image of the sample fluorescence — and, correspondingly, the area of the diamond crystal under optical excitation — is presented in Fig. 2b: Rather than illuminating the sample uniformly, we focus the laser beam down to a 300-µm-wide spot; we later show this geometry is instrumental in attaining the highest spin pumping efficiency. The red trace in Fig. 2c shows the $^{13}$C NMR spectrum at 9.4 T after optical illumination for an interval $t_L = 10$ s at 51.7 mT (approximately coincident with the $^{13}$C spin-lattice relaxation time $T_1^{(C)} \approx 6$ s at this field); the number of repetitions is only 4 and the total experimental time amounts to about 30 s. Comparison with the thermal spectrum at high field (central blue trace, 24 repetitions) indicates a ~30-fold signal gain. Taking into account the fractional sample volume under illumination (~1% of the total), we conclude that the optically-pumped $^{13}$C polarization $P_C$ amounts to about 3%, i.e., the equivalent of a 3000-fold enhancement over the thermal polarization at 9.4 T. Note that given the long $^{13}$C spin-lattice relaxation time $T_1$ at high field (of order 15 minutes), the time required to obtain the same signal-to-noise ratio (SNR) without optical pumping would amount to ~600 hours.

Despite the large signal enhancement, identifying the mechanisms responsible for the electron-nuclear spin polarization transfer is not immediate. ODMR experiments indicate that the ¹³C spins most strongly coupled to an NV⁻ — i.e., within the first or second atomic shells — polarize efficiently near ~51.2 mT, where the NV⁻ excited triplet state experiences a level anti-crossing² (LAC). The mechanism resembles that invoked for the dynamic polarization of the ¹⁴N spin of the nitrogen host¹¹, namely, mixing between the spin states in the NV⁻ excited triplet produces rapid 'flip-flops' between electron and nuclear spins. Therefore, in the presence of continued laser excitation, spin-selective triplet-singlet intersystem crossing drives the electron-nuclear spin system jointly into a state of net polarization. In this picture, alignment of 'bulk' ¹³C nuclei — i.e., nuclei with negligible hyperfine interaction — takes place via spin diffusion from NVs³. Note that this process is inherently slow because, besides the low ¹³C concentration and weak homonuclear couplings, the energy mismatch between carbons with differing hyperfine couplings inhibits spin-exchange (i.e., ¹³C–¹³C flip-flops) except during the times the NV⁻ spin is in the $m_S = 0$ state (i.e., when the hyperfine coupling in the ground state effectively vanishes).

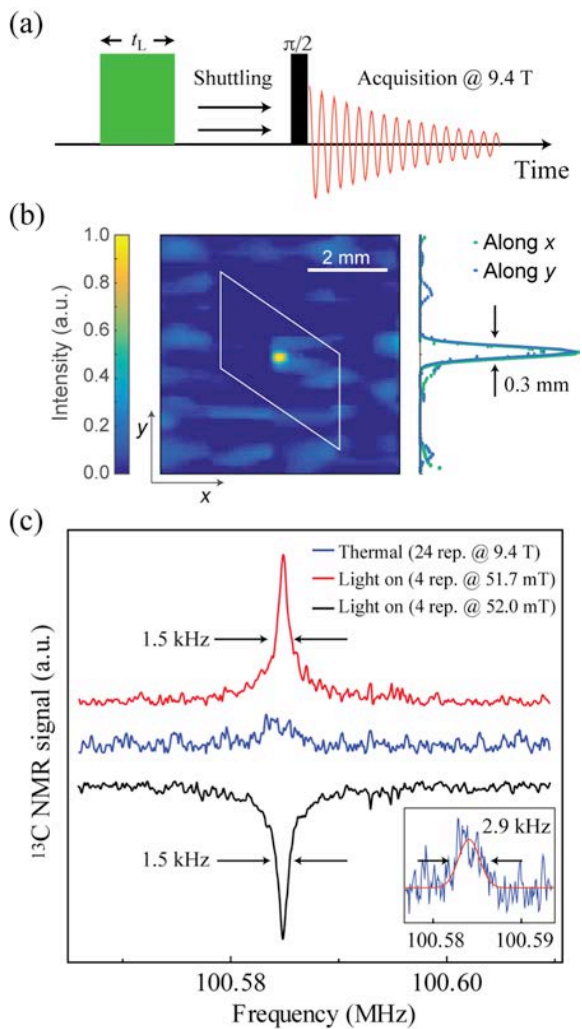

**Fig. 2:** (a) Pulse sequence. After green illumination for a time $t_L$ in a magnetic field $B_P$ we shuttle the probe head to the superconducting magnet for detection at $B_0 = 9.4$ T. (b) Using a pair of lenses we focus the laser beam to a 0.3 mm diameter spot on the 3.2×3.2×0.3 mm³ diamond. Using a camera and an optical filter, we record the fluorescence image shown in the picture. The superimposed white square is a contour of the crystal. (c) ¹³C NMR spectra after optical pumping at 51.7 mT (red trace) and 52.0 mT (black trace); in either case, the total number of repeats is 4, the illumination time is $t_L = 10$ s, the laser power is $P_L = 1$ W, and the magnetic field $B_P$ is nearly parallel to the diamond [111] axis. For reference, the blue trace shows the thermal spectrum at 9.4 T for a total of 24 repeats; the wait time between repeats is 1200 s. All spectra have been displaced vertically for clarity.

Unfortunately, this intuitive scenario is difficult to reconcile with our observations. For example, we find that the optically induced ¹³C magnetization reverses upon a slight change in the applied pumping field $B_P$. An illustration is shown in Fig. 2c displaying a full sign inversion at $B_P = 52.0$ mT (black trace), only 0.3 mT away from the conditions discussed above (red trace). Given the large hyperfine couplings with proximal ¹³C spins (of order ~5 mT for first shell nuclei¹²⁻¹⁵), such extreme sensitivity hints at co-existing but countering spin polarization channels (e.g., provided by carbons with opposing hyperfine couplings), and hence to an irregular dependence of the ¹³C NMR signal on the magnitude of the applied magnetic field.

Instead, a systematic characterization of the optically-induced ¹³C spin polarization as a function of the applied pumping field shows this is not the case (Fig. 3a). We find a fairly symmetric response featuring a collection of alternating maxima and minima of comparable amplitude. Grouping consecutive extrema into pairs, we identify an odd number of magnetic fields $B_P^{(j)}$, $j = -2 \ldots 2$ (black dashed lines in Fig. 3) around which the polarization transfer is most efficient. Optimum enhancement is attained near the center (i.e., around $B_P^{(0)}$) but the pattern also exhibits two prominent, nearly symmetric shoulders of comparable amplitude near $B_P^{(\pm 1)}$. Additional polarization extrema of lower amplitude are clearly present near $B_P^{(\pm 2)}$ — notably half way between $B_P^{(0)}$ and $B_P^{(\pm 1)}$— and to the right (left) of $B_P^{(+1)}$ ($B_P^{(-1)}$).

To interpret this dependence we resort to mechanisms that build on the presence of substitutional nitrogen, by large the most abundant impurity in our Type 1b diamond (see Fig. 3b). In the neutral charge state, nitrogen is paramagnetic (spin number $S' = 1/2$) and known to cross-relax with NV⁻.¹⁶ In particular, efficient cross-relaxation can exist near ~51 mT where the $|m_S' = +1/2\rangle \leftrightarrow |m_S' = -1/2\rangle$ electronic spin transition of the nitrogen nearly matches the $|m_S = 0\rangle \leftrightarrow |m_S = -1\rangle$ ground state transition of the NV. Unlike NVs, however, P1 centers have a strong hyperfine coupling with the

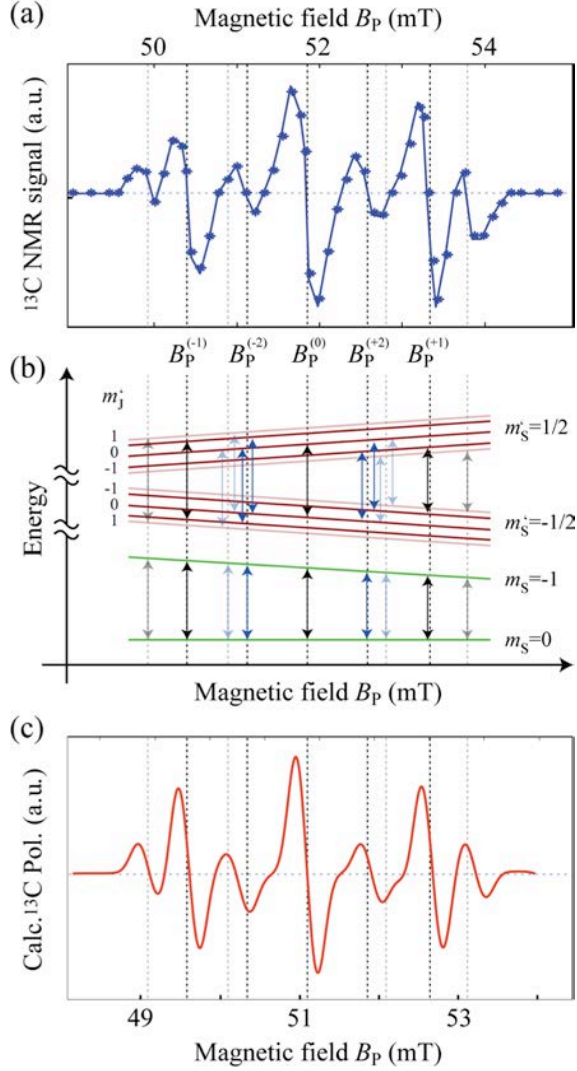

**Fig. 3:** (a) Amplitude of the $^{13}$C OPNMR signal (upper half) and NV$^-$ fluorescence (lower half) as a function of the applied polarization field $B_P$. Solid and dotted lines are guides to the eye. The experimental conditions are those of Fig. 2. (b) Level diagram for the NV and P1 centers (green and red, respectively) as a function of the applied magnetic field $B_P$ in the vicinity of 50 mT. The dashed lines indicates the fields where the frequency of the NV$^-$ $|m_S = 0\rangle \leftrightarrow |m_S = -1\rangle$ electronic transition matches one of the hyperfine-shifted $|m'_S = -1/2\rangle \leftrightarrow |m'_S = +1/2\rangle$ transitions of the P1 center; $m'_J = 0, \pm 1$ denotes the quantum projection of the $^{14}$N nuclear spin at the P1 center along the [111] axis. The diagram neglects the $^{14}$N-induced hyperfine splitting at the NV$^-$ site, much smaller than for the P1 center. (c) Calculated $^{13}$C polarization pattern (see text) in the case where the NV axis is perfectly aligned with the external magnetic field.

nuclear spin of the host nitrogen (of order ~100 MHz). Therefore, energy matching is predicted at three different magnetic fields — here coincident with $B_P^{(0)}$ and $B_P^{(\pm 1)}$ — depending on the quantum projection of the P1 nuclear spin state, $m'_J = 0, \pm 1$ ($^{14}$N has spin $J' = 1$ and is 99.6% abundant). Cross polarization of the $^{13}$C nuclei takes place around each of these fields via a three-spin 'flip'; carbons polarize in one direction or the other depending on the sign of the mismatch between the NV and the P1 transition frequencies. The result is a dispersive-like response centered at each matching magnetic field, whose width is defined by the probability governing the distribution of hyperfine couplings (along with the NV's own hyperfine coupling with its host $^{14}$N, see below).

The observed dependence is more complex, however, because P1 centers have trigonal symmetry (i.e., the nitrogen moves slightly away from the lattice site towards one of the four nearest carbons). Thus, for a magnetic field parallel to one of the C–C bonds, P1s split into two subsets, each featuring a characteristic hyperfine coupling (faded traces in the schematics of Fig. 3b). $^{13}$C spin polarization to the right (left) of $B_P^{(+1)}$ ($B_P^{(-1)}$) stems from flips mediated by P1s parallel to the [111] axis (featuring the largest hyperfine coupling constant). Because the number of P1 centers oriented along any of the three remaining bonds (forming an angle of $70.5^0$ with the [111] axis) is three times larger, we expect greater $^{13}$C NMR signals around $B_P^{(\pm 1)}$. Further, since both P1 subsets contribute to the nuclear spin polarization around $B_P^{(0)}$, we anticipate relative amplitudes of the form 4:3:1 for the NMR signal around the central and satellite transitions, in qualitative agreement with our observations.

The appearance of $^{13}$C polarization around $B_P^{(\pm 2)}$ can be understood along similar lines: In this case the spin transfer is mediated simultaneously by the electron and nuclear spin of the P1 center, both of which 'flip' to match the energy necessary for an NV/$^{13}$C 'flop'. As is evident from the schematic in Fig. 3b, such processes must take place around magnetic fields midway between $B_P^{(0)}$ and $B_P^{(+1)}$ (blue arrows). This mechanism should be equally active in P1 centers of all four orientations, but because the resulting polarization is comparatively lower than in the case considered before, only the greater contribution from misaligned P1s is apparent in the experimental pattern. The end result of a numerical model that takes into account all these interactions shows excellent agreement with our observations (Fig. 3c).

The derivation of the polarization pattern above is worth considering in detail. We start with the model Hamiltonian

$$H = H_{\text{NV}} + H_{\text{P1}} + H_{\text{NV-P1}} + H_C + H_N + H_{N'} + \\ + \mathbf{S} \cdot \mathbf{A_C} \cdot \mathbf{I} + \mathbf{S} \cdot \mathbf{A_N} \cdot \mathbf{J} + \mathbf{S}' \cdot \mathbf{A_{N'}} \cdot \mathbf{J}', \quad (1)$$

where $\mathbf{A_C}$ is the hyperfine coupling tensor between the NV$^-$ electronic spin $\mathbf{S}$ and a $^{13}$C spin $\mathbf{I}$, and $\mathbf{A_N}$ ($\mathbf{A_{N'}}$) is the hyperfine coupling tensor between the NV$^-$ (P1) electronic spin and the $^{14}$N nuclear spin $\mathbf{J}$ ($\mathbf{J}'$) of the nitrogen host. $H_{\text{NV}}$ denotes the NV$^-$ Hamiltonian expressed as the sum of the crystal field $\Delta S_z^2$ and Zeeman interaction $-\gamma_e B_P S_z$, $H_{P1} = -\gamma_e B_P S_z'$ represents the P1 Zeeman energy, and $H_{NV-P1}$ is the dipolar coupling between the NV and P1 centers. Finally, $H_C = -\gamma_C B_P I_z$ is the $^{13}$C Zeeman Hamiltonian, and $H_{N'} = -\gamma_N B_P J_z' + Q' J_z'^2$ is the $^{14}$N Hamiltonian at the P1, including Zeeman and quadrupolar terms; a similar expression holds for $H_N$, the Hamiltonian of the $^{14}$N at the NV center, $\gamma_C$, $\gamma_N$ respectively denote the $^{13}$C and $^{14}$N gyromagnetic ratios, and $Q'$ ($Q$) is the quadrupole constant for the $^{14}$N host in a P1 (NV$^-$) center[17]. In the above expressions we assume the NV center is aligned with the magnetic field, and make the NV$^-$ and P1 gyromagnetic ratios equal to the electron gyromagnetic ratio $\gamma_e$.

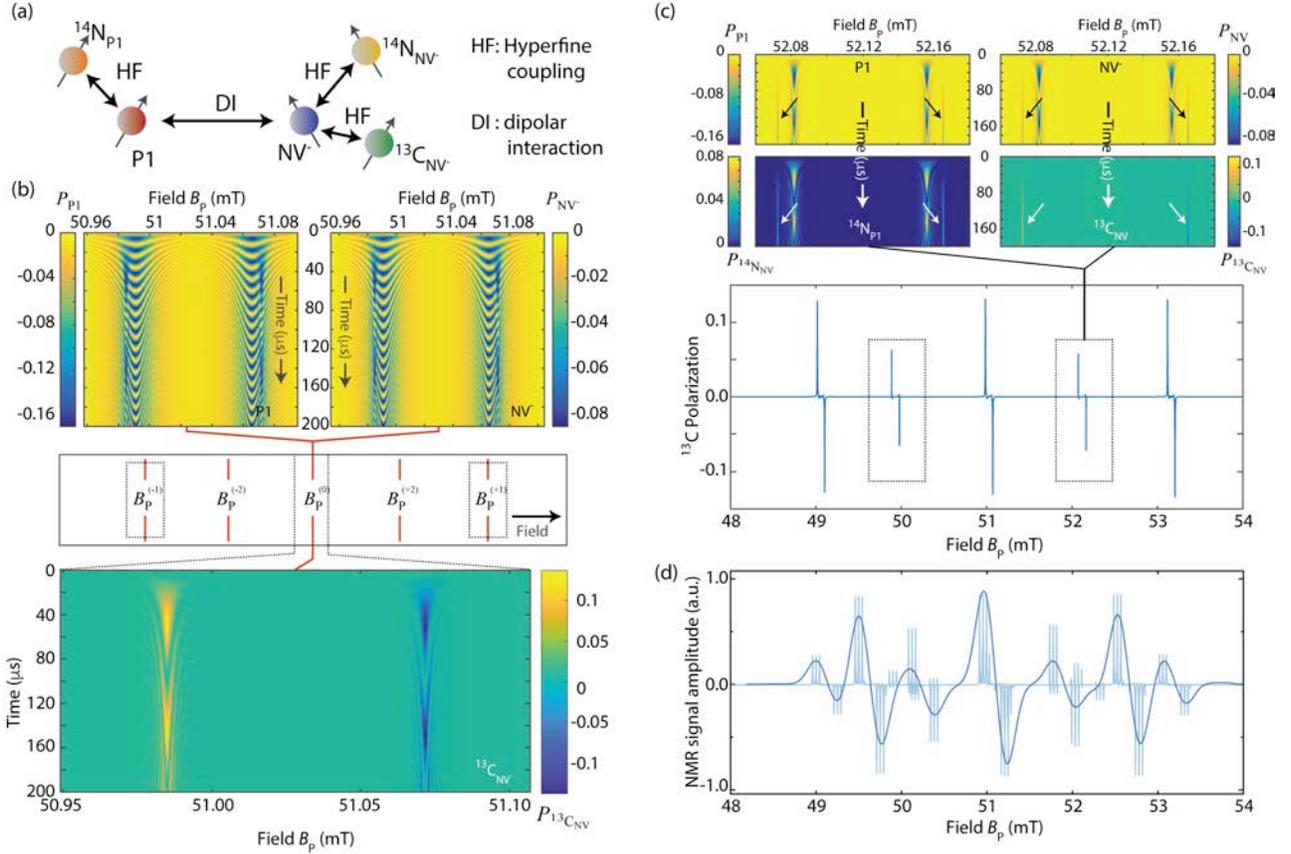

**Fig. 4:** (a) We consider a model of a single $^{13}$C spin coupled to an NV$^-$ center, itself interacting with its host $^{14}$N spin and a neighboring P1 center; the latter experiences a hyperfine coupling with its own host $^{14}$N. (b) (Top) Calculated unitary time-evolution of the P1 and NV$^-$ electronic spins (top left and right, respectively) and the $^{13}$C nuclear spin (bottom) in a vicinity of $B_P^{(0)}$; a similar response is obtained near $B_P^{(\pm 1)}$ as indicated by the dashed lines. In all cases, the system starts from a state where the NV$^-$ spin is in $m_S = 0$, its host $^{14}$N nuclear spin is in $m_J = 0$, and all other spins are in thermal equilibrium. (c) (Bottom) Projected amplitude of the $^{13}$C spin polarization as a function of $B_P$ assuming the NV and P1 axes coincide with the direction of the field. (Top) Time evolution of the P1 and its $^{14}$N host (left) and the NV$^-$ and coupled $^{13}$C spin (right) polarizations in a vicinity $B_P^{(+2)}$. Arrows highlight the four-spin dynamics. A similar response is found near $B_P^{(-2)}$ (dashed rectangle). (d) Modeled $^{13}$C NMR signal as a function of $B_P$ starting from a state where all spins (including the $^{14}$N at the NV) are in thermal equilibrium (fainter blue trace). This model also takes into account cross-relaxation with P1 centers both parallel and non-parallel with the field (see main text). The starker solid line is a convolution of the calculated line pattern with a 0.1 mT Gaussian. The NV$^-$–$^{13}$C coupling is 4 MHz in (b, c) and 13 MHz in (d); in all three cases the NV$^-$– P1 coupling is 500 kHz.

Evolving from a density matrix $\rho_0$ where the NV$^-$ and nuclear spin of its nitrogen host are in the $m_S = 0$ and $m_J = 0$ states, respectively, and all other spins (i.e., the $^{13}$C spin and the electron and nuclear spins of the P1, see Fig. 4a) are in thermal equilibrium, we first calculate the net P1 and NV$^-$ spin polarizations. These are respectively defined as $P_S(t) \equiv Tr\{S_z \rho(t)\}$ and $P_{S'}(t) \equiv 2Tr\{S_z' \rho(t)\}$, where $\rho(t) = U\rho_0 U^\dagger$ and $U = exp(-iHt/\hbar)$ denotes the time evolution operator at time $t$. For both, we find a similarly complex response, symmetrically split about $B_P^{(0)}$, a consequence of the hyperfine coupling with the $^{13}$C spin (upper half of Fig. 4b); we observe an analogous behavior around $B_P^{(\pm 1)}$, not shown for simplicity. Besides the fast oscillations, both electronic spins undergo a slower dynamics that correlates with the $^{13}$C spin evolution (compare with lower half of Fig. 4b), indicative of a three-spin process. Unlike the P1 or the NV$^-$ spins, however, the latter polarizes positively or negatively, depending on the shift relative to $B_P^{(0)}$ (or $B_P^{(\pm 1)}$), as observed experimentally. We note that despite the strong hyperfine coupling with the paramagnetic center, the $^{14}$N spin at the P1 site remains unpolarized near these fields (not shown).

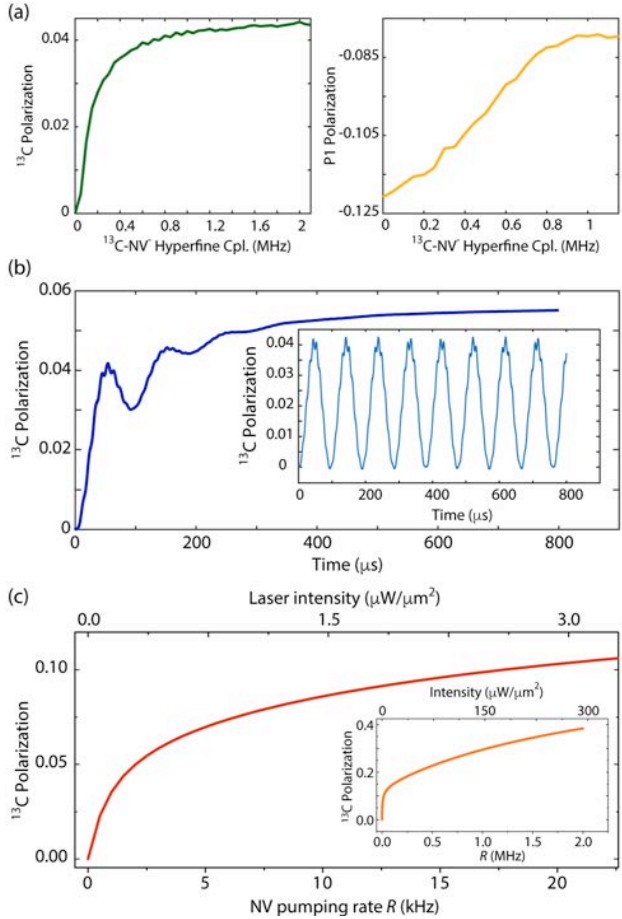

**Fig. 5:** (a) Calculated absolute amplitude of the $^{13}$C and P1 spin polarization near $B_P^{(0)}$ (left and right, respectively) as a function of the $^{13}$C hyperfine coupling with the NV$^-$. In these calculations we assume that both host $^{14}$N nuclear spins are in thermal equilibrium (i.e., none is polarized). (b) Calculated time evolution of the $^{13}$C polarization in the presence of continuous illumination assuming an NV spin pumping rate $R$=9 kHz. For reference, the insert shows the result of a unitary evolution in the dark. (c) Calculated steady-state $^{13}$C spin polarization as a function of the NV spin pumping rate for $^{13}$C, NV, and P1 spin lattice relaxation times $T_1^{(C)} = 10$ s, $T_1^{(NV)} = 1$ ms, and $T_1^{(P1)} = 100$ μs respectively; the spin-lattice relaxation times of both $^{14}$N hosts is $T_1^{(N,N')} = 100$ ms. The insert shows the same but for a larger pumping rate range.

The above dynamics fundamentally changes near $B_P^{(\pm 2)}$: As seen in the zoomed plots on the upper half of Fig. 4c, the $^{13}$C evolution correlates well with that of the NV$^-$ spin and the electron and nuclear spins at the P1 center, implying that the mechanism of $^{13}$C spin polarization at these fields is the result of a double electron-nuclear process[18]. In other words, exactly at $B_P^{(\pm 2)}$ energy matching between the NV$^-$ and P1 transitions is reached by an accompanying flip of the P1 nitrogen nuclear spin (see Fig. 3b); $^{13}$C spins polarize in one direction or the other as small shifts in the value of $B_P$ introduce a slight energy difference. A projection of the $^{13}$C polarization amplitude throughout the full range of hyperfine-shifted P1 transitions is presented in the lower half of Fig. 4c, exposing all five polarization fields for a spin set with collinear P1 and NV centers.

Fig. 4d shows an extension of the model that also includes contributions from spin clusters where the P1 and NV centers are oriented along different crystallographic axes (a scenario three times more likely than the alternative considered above). To take into account the statistical distribution of $^{13}$C hyperfine couplings, we convolve the calculated field dependence of the NMR signal amplitude with a Gaussian, whose width (~0.1 mT) is extracted from the measured data (Fig. 3a). The $^{13}$C NMR signal pattern that emerges is in very good agreement with our experimental results. Worth highlighting in particular are the relative amplitudes near $B_P^{(\pm 2)}$ — about half of those attained around $B_P^{(0,\pm 1)}$ — which reasonably reproduces our observations. Importantly, the broadening imposed above also takes into account the hyperfine shift of the NV nitrogen nuclear spin (±2.3 MHz, or the equivalent of ~±0.08 mT), here assumed in thermal equilibrium. Since the splitting between negative and positive $^{13}$C polarization scales with the hyperfine coupling to the NV (see below), we surmise that only those carbons where the coupling is sufficiently strong (i.e., larger than ~2 MHz) constructively contribute to the $^{13}$C polarization buildup. More weakly coupled carbons, however, may also play a role if the NV$^-$ axis is precisely aligned with the external magnetic field. In this limit, the nuclear spin of the $^{14}$N host polarizes via the excited state level anti-crossing[11], arguably eliminating the hyperfine broadening and correspondingly leading to narrower features in the observed field-dependence of the NMR signal amplitude. Since only a few-degree misalignment is sufficient to inhibit this process, this condition proved difficult to attain in the present experiment (see below).

Given the near-unity spin initialization possible in NV$^-$, an interesting problem is the (absolute) level of $^{13}$C polarization that can be pumped at a given magnetic field (for a given $^{13}$C spin-lattice relaxation time). To answer this question we first calculate the maximum $^{13}$C (and P1) spin polarization after unitary evolution in a vicinity of $B_P^{(0)}$ for variable $^{13}$C–NV$^-$ hyperfine coupling. As shown in Fig. 5a, we find a sensitive response that ranges from no polarization for weakly-coupled $^{13}$C spins to an optimum of 4% for 2 MHz hyperfine couplings or stronger. The P1 center, on the other hand, gradually reduces its polarization from ~11% — nearly ideal if one takes into account the thermal mixing from $^{14}$N-induced hyperfine splittings of the NV and P1 centers — to about 8.5 %. We surmise, therefore, that the notion of a P1-mediated 'flip-flop' must be viewed as a crude simplification.

For hyperfine couplings greater than 2 MHz, the limit $^{13}$C spin polarization emerges from the interplay between the carbon spin-lattice relaxation time $T_1^{(C)}$, and the NV spin re-pumping rate $R$, itself a function of the laser intensity and NV spin-lattice relaxation $T_1^{(NV)}$; additional parameters of importance are the spin-lattice relaxation times of the P1 spin $T_1^{(P1)}$ and nitrogen hosts (respectively $T_1^{(N)}$ and $T_1^{(N')}$ for the $^{14}$N spins at the NV and P1 sites). One first illustration is shown in Fig. 5b where we plot the result from quantum mechanical calculations modified to take into account the effect of continuous optical excitation during evolution (Appendix A). For $R = 9$ kHz the $^{13}$C polarization plateaus near the limit value in a unitary transfer. Higher nuclear spin polarization — corresponding to $^{13}$C 'spin pumping' — can be attained at higher NV spin pumping rates. Fig. 5c presents the calculated steady state $^{13}$C spin polarization as a function of $R$ assuming a nuclear spin-lattice relaxation time $T_1^{(C)} = 10$ s. $^{13}$C spin polarization exceeding 10% is predicted for laser intensities as low as ~3 μW/μm$^2$ (corresponding to $R$~20 kHz). We calculate higher $^{13}$C polarization — exceeding 20% — for large laser intensities (>70 μW/μm$^2$ corresponding to $R \geq 0.5$ MHz) though this limit is impractical in experiments with macroscopic samples as other deleterious effects (e.g., thermal) are likely to prove dominant.

An immediate outcome of a P1-mediated dynamics is the prediction of robust spin polarization transfer in less than 'ideal' geometries, i.e., where the NV and magnetic field axes form a non-zero angle $\theta$. Fig. 6a shows the results from observations as a function of the magnetic field amplitude $B_P$ for different alignment conditions. Consistent with a P1-enabled transfer, we attain a similar response of the NMR signal on the external field with comparable levels of peak $^{13}$C spin polarization for angles $\theta$ as big as 17 degrees. Given the dependence of the NV$^-$ transition frequencies on the field orientation, the polarization extrema cluster around $\theta$-dependent fields $B_P^{(j)}$, each highlighting as before a different energy matching condition. We attain excellent agreement between the calculated and measured values of $B_P^{(0)}$ at different angles $\theta$ over a wide range (Fig. 6b), hence confirming the important role P1 centers play in the observed dynamics.

Since the angle between any two possible NV axes is 70.5 degrees, the maximum misalignment $\theta_m$ amounts to ~35 degrees, only about twice the range of angles covered herein. Because energy-matching fields can be identified for the full range of NV orientations (see insert in Fig. 6b), it would be reasonable to anticipate optical pumping of $^{13}$C polarization over a broad range of magnetic fields, from ~50 to ~85 mT. The latter, of course, depends on the degree of NV$^-$ spin polarization, known to decrease when the magnetic field is misaligned. Therefore, extending the present study to NV orientations closer to $\theta_m$ should prove useful to experimentally determine the degree of level mixing in the excited NV$^-$ state, the factor ultimately governing the nuclear spin pumping efficiency.

Another interesting facet in the present experiments is the illumination geometry, unexpectedly found to have an intriguing impact on the $^{13}$C spin polarization. Fig. 7a compares three different regimes, with the laser beam gradually exciting a smaller fraction of the sample. The measured $^{13}$C NMR signal amplitude as a function of the input laser power for each geometry is shown in Fig. 7b: In all cases we find an exponential growth featuring comparable saturation laser powers and limit amplitudes, a puzzling response given the large differences between the excited sample volumes (up to a factor ~40 between the two limit cases). To compare the observed and expected NMR amplitudes on a more quantitative basis, we first use ODMR to measure the NV$^-$ spin polarization $P_{\mathrm{NV}}$ in this sample as a function of laser intensity. For an arbitrary but fixed laser power $W$, we can then obtain a relative measure $s_j$ of the expected $^{13}$C signal amplitude through the formula

$$s_j(W) = \int_A da\ U_j(\mathbf{r})\, P_C\big(U_j(\mathbf{r})\big)\, P_{\mathrm{NV}}\big(U_j(\mathbf{r})\big), \quad (2)$$

where $U_j(x,y)$ denotes the laser power profile for a given beam geometry $j = i, ii, iii$ (the index $j$ growing with the beam radius), $P_{\mathrm{NV}}$ denotes the NV$^-$ spin polarization as derived from the ODMR contrast, $P_C$ is the $^{13}$C polarization as calculated in Fig. 5c (where the NV is assumed polarized), $\mathbf{r} = (x,y)$ is the position vector on the illuminated area, and the integral extends over the sample surface $A$. In Eq. 2 we normalize each intensity profile $U_j(\mathbf{r})$ so that $W = \int_A da\ U_j(\mathbf{r})$ and assume the magnetic field direction coincides with one of the four possible NV axes.

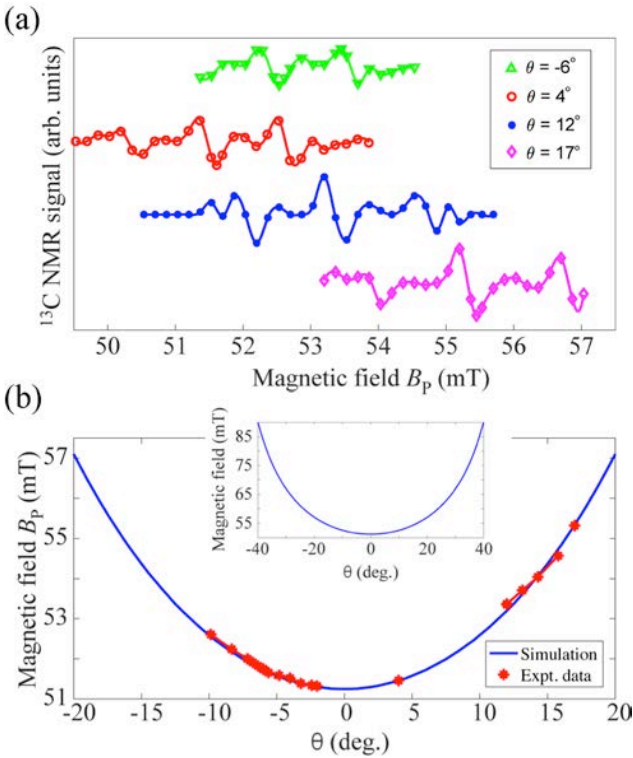

**Fig. 6:** (a) $^{13}$C NMR signal amplitude as a function of the polarization field $B_P$ for four different orientations $\theta$ of the [111] axis relative to the field direction. Comparable levels of $^{13}$C spin polarization (2-4%) can be attained in all cases. (b) Measured $B_P^{(0)}$ as a function of $\theta$ (solid red circles). The solid blue line is the calculated value, shown in the upper insert for a larger range of angles. Polarization transfer to the $^{13}$C should therefore be possible for all NV orientations. In (a) traces have been displaced vertically for clarity; solid lines are guides to the eye.

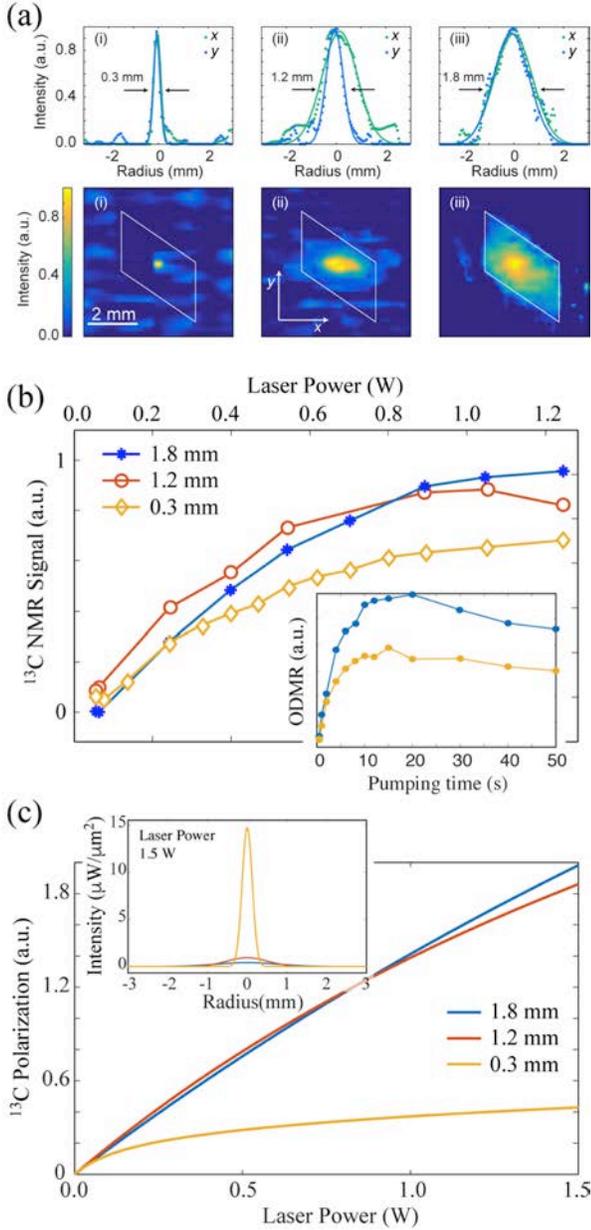

**Fig. 7:** (a) We compare the system response under varying illumination geometries. Fluorescence images and the corresponding cross sections are shown in the lower and upper rows, respectively. In (i) through (iii), the effective beam radius is 0.3 mm, 1.2 mm, and 1.8 mm, respectively. The white parallelogram marks the contour of the diamond crystal. (b) (Main) $^{13}$C NMR signal as a function of laser power for all three geometries. (Insert) $^{13}$C NMR signal amplitude as a function of illumination time; the laser power is 1.2 W. In both cases the magnetic field is 51.7 mT (resulting in maximum NMR signal, see Fig. 3a) and its direction nearly coincides with one of the NV axes. (c) Calculated signal amplitudes as derived from Eq. (2) in the main text. (Insert) Calculated intensity profiles assuming the total beam power is 1.5 W.

A vis-à-vis comparison between the calculated and experimental signal amplitudes (Figs. 7b and 7c, respectively) exposes substantial differences. For example, for laser powers around 1 W, we find $s_{iii} \sim 5 s_i$, substantially larger than observed. Perhaps more intriguingly, the comparable saturation laser powers observed in all geometries is at odds with the calculated response, increasing more slowly for the larger beams. The latter is, of course, a manifestation of the relatively low laser intensities in these geometries, and hence the less-than-optimum NV$^-$ spin polarization being optically pumped. That greater laser powers do not lead to greater observable NMR signals is therefore indicative of underlying dynamics not properly accounted for by the present model.

One possibility to take into account is sample heating. While diamond is transparent to optical illumination, the process required to produce high NV concentrations also leads to other, unwanted point defects lending the crystal a noticeable greyish color. Given the enhanced optical absorption and reduced thermal conductivity typical in this type of diamond, a substantial rise in sample temperature is likely (despite the palliative effect of the sapphire substrate). Because the NV$^-$ energy level separation is a sensitive function of temperature[19] (the thermal coefficient near room temperature amounts to ~76 kHz/K), sample heating alters the energy matching conditions during the multi-second illumination interval. This problem worsens with greater laser powers meaning that the maximum absolute polarization emerges from a complex interplay between spin optical pumping, heat production, and thermal dissipation.

An indication on the impact of temperature on the system dynamics is presented in the insert to Fig. 7b, where we plot the $^{13}$C NMR signal amplitude for varying optical pumping intervals. We find similar trends both for beam geometries (i) and (iii), namely, the NMR signal grows exponentially to reach a maximum after ~10 s of illumination, but subsequently reverses its trend to exhibit a slow decay. This behavior is consistent with the ideas above, i.e., longer light exposure shifts the energy matching condition away from the optimum near 52 mT (Fig. 3a), and thus leads to a progressive reduction of the nuclear spin polarization. Using as a reference the magnetic field dependence of the optically pumped NMR signal in Fig. 3a — reversing sign upon a change of order 0.2-0.3 mT — we conclude temperature changes of order 20-40 K would be sufficient to have a sizable impact.

The considerations above assume, of course, that the fraction of NVs in the negatively charged state remains constant throughout the multi-second long intervals of continuous illumination common in optically-pumped NMR experiments. This is, however, not the case: Initial fluorescence microscopy observations in this sample (not shown here for brevity) indicate a slow but sustained conversion from $NV^-$ to $NV^0$ under continuous laser light. Different from two-photon charge conversion [20], the mechanism underlying this process — observed to rapidly reverse in the dark — is presently unknown, but we hypothesize it is driven by forward electron tunneling from the optically excited $NV^-$ to neighboring $N^+$ ions. This form of photo-induced $NV^-$ depletion — which is intensity dependent — can help explain the large disagreement between the measured and calculated NMR signals in Fig. 7. By the same token, fluctuating magnetic fields arising from moving charges — including those from P1 center photo-induced ionization — can also accelerate nuclear spin relaxation[21]. Combined with $NV^-$ depletion, this mechanism may lead to the gradual depolarization of $^{13}C$ spins shown in the insert to Fig. 7b. Improved hardware — for example, in the form of actively cooled sample holders — as well as extended optical spectroscopy and magnetic resonance experiments will be necessary to separate the impact of thermal heating and charge-related effects on the system response.

### III. CONCLUSIONS

In summary, we attain efficient optical pumping of bulk $^{13}C$ spin polarization in diamond of up to ~3% at room temperature. As evidenced by the characteristic response with the applied magnetic field, the dynamics is governed by cross-relaxation with the electron (and nuclear) spin(s) of adjacent P1 centers, coupled to the NV via dipolar interactions. The transfer process is energy conserving, i.e., nuclear spins polarize positively or negatively depending on the difference between the $NV^-$ and P1 resonance frequencies at a given field. This mechanism remains efficient over a broad range of crystal orientations, provided the amplitude of the magnetic field is adjusted to nearly match the $NV^-$ and P1 resonances. Since the level of $^{13}C$ polarization (2-4%) remains unchanged for all probed crystal orientations — about half the possible range — we surmise the impact of level mixing on the NV spin polarization is low. On the other hand, laser-induced heating appears to have a noticeable, deleterious effect, though additional work will be needed to disentangle the interplay between spin pumping, temperature drifts, and $NV^-$ and P1 photo-ionization.

Along the same lines, the role played by the $NV^-$ excited state LAC in the observed polarization buildup warrants further examination: $^{13}C$ spins strongly coupled to an $NV^-$ center are thought to polarize efficiently via state mixing at the excited state LAC[11], but the excellent agreement with the P1-driven model observed herein suggests this mechanism is comparatively inefficient. The reason for this stark difference is presently unclear but one possible scenario — not considered thus far — involves the cross-polarization of $^{13}C$ spins coupled to P1 centers (rather than to the NV). Given the comparatively larger concentration of nitrogen, nuclear spin polarization is arguably seeded more uniformly in a P1-enabled process, hence accelerating the polarization growth by effectively lowering the wait time inherent to a spin-diffusion-dominated dynamics. Preliminary calculations (see Appendix B) show that P1-coupled carbons do indeed polarize (although the time scale is slower than the case we considered in the main text, see Fig. 4a). On the other hand, the calculated dependence on the applied field shows only moderate agreement with our observations, suggesting that a simple change in the Hamiltonian from $NV^-$- to P1-coupled carbons is insufficient to explain the dominance of cross-polarization over LAC-mediated spin pumping. Future work should address this problem by considering more complex multi-spin clusters (e.g., where carbons cross-relax via interaction with two coupled P1 centers) and through experiments combining optical and microwave excitation.

### IV. ACKNOWLEDGEMENTS

We thank Neil Manson and Marcus Doherty for kindly providing the sample for these experiments. We also thank Friedemann Reinhard, Alejandro Bussandri, Carlos Sanrame, and Vinod Menon for fruitful discussions and assistance with some of the experiments. D.P., R.K.K., H.H.W., A.A., S.D., P.R.Z., and C.A.M. acknowledge support from the National Science Foundation through grants NSF-1309640 and NSF-1401632, and from Research Corporation for Science Advancement through a FRED Award; they also acknowledge access to the facilities and research infrastructure of the NSF CREST Center IDEALS, grant number NSF-HRD-1547830.

### APPENDIX A

Unitary spin dynamics is evaluated by exact diagonalization of the Hamiltonian in Eq. (1). In order to include the effects of light and relaxation processes ($T_1$-driven depolarization), we resort to a quantum master equation (QME) with the corresponding Lindblad terms. Here, the QME was solved using the QuTiP Python library.[22] In general, the QME that describes the evolution of the density matrix $\rho$ is given by

$$\dot{\rho} = -\frac{i}{\hbar}[\hat{H},\rho] + \sum_{k=1}^{K} \hat{c}_k \rho \hat{c}_k^\dagger - \frac{1}{2}\hat{c}_k\hat{c}_k^\dagger \rho - \frac{1}{2}\rho \hat{c}_k\hat{c}_k^\dagger, \quad (A.1)$$

where each term $k$ in the summation can be associated to a different dissipative process. For example, in the case of laser illumination, i.e. the optical pumping of the NV$^-$ spin, there are two simultaneous pathways contributing to spin initialization: The transfer $k'$ of population from $m_S = +1$ to $m_S = 0$ and the transfer $k''$ from $m_S = -1$ to $m_S = 0$. These are respectively represented by two Lindblad operators that act only in the NV Hilbert space as $\hat{c}_{k'} = \sqrt{R}|0\rangle\langle+1|$ and $\hat{c}_{k''} = \sqrt{R}|0\rangle\langle-1|$; here, $R$ stands for the NV pumping rate.

To take into account the different spin-lattice relaxation times $T_1$ governing each spin species, we introduce a two-way population transfer $k'''$ between every pair of spin states, $\hat{c}_{k'''}^{(m,n)} = \sqrt{\Gamma_1}|m\rangle\langle n|$ and $\hat{c}_{k'''}^{(n,m)} = \sqrt{\Gamma_1}|n\rangle\langle m|$. Here, $\Gamma_1 \equiv 1/T_1$ for that particular spin. This means that the summation in the second term of the QME includes six relaxation processes acting on the NV, two acting on the $^{13}$C, six on each of the host nitrogen nuclear spins, and two processes on the P1 spin.

Due to the broad range of time-scales involved in the dynamics, the analysis of the long time asymptotics of the polarization can be computationally demanding. In order to avoid the explicit evaluation of the time-dependence for every pumping rate $R$, we calculate the stationary state $\rho_\infty$ that the system reaches at sufficiently long times. This state is obtained by solving the algebraic equation,

$$\dot{\rho}_\infty = 0 = -\frac{i}{\hbar}[\hat{H},\rho_\infty] + \sum_{k=1}^{K} \hat{c}_k \rho_\infty \hat{c}_k^\dagger \\ -\frac{1}{2}\hat{c}_k\hat{c}_k^\dagger \rho_\infty - \frac{1}{2}\rho_\infty \hat{c}_k\hat{c}_k^\dagger. \quad (A.2)$$

By including not only the effects of light but also the spin-lattice relaxation for all spins in the system, $\rho_\infty$ turns out to be a function of $R$ and all the $T_1$ time-scales involved. Then, for a given $\rho_\infty$ we calculate the steady state expectation value of the $^{13}$C polarization, as shown in Fig. 5 (c) of the main text.

## APPENDIX B

In an alternate nuclear spin polarization mechanism briefly considered in the text, cross relaxation between NV$^-$ and P1 spins takes place with the assistance of $^{13}$C spins coupled to the P1 center rather than to the NV. In order to explore such a scenario, we modify the model Hamiltonian to yield,

$$H = H_{NV} + H_{P1} + H_{NV-P1} + H_{C'} + H_N + H_{N'} \\ + \boldsymbol{S'} \cdot \boldsymbol{A}_{C'} \cdot \boldsymbol{I'} + \boldsymbol{S} \cdot \boldsymbol{A}_N \cdot \boldsymbol{J} + \boldsymbol{S'} \cdot \boldsymbol{A}_{N'} \cdot \boldsymbol{J'}. \quad (B.1)$$

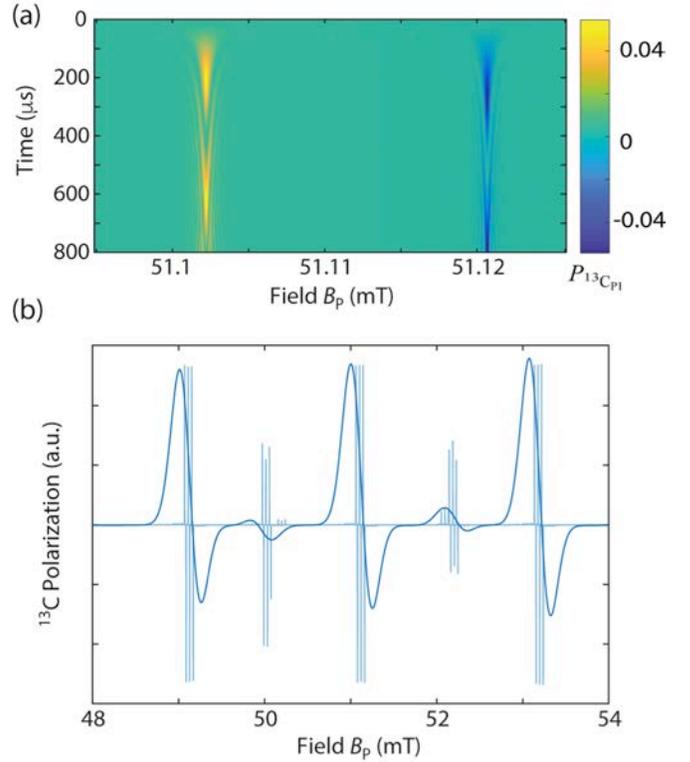

**Supplemental Fig. 1:** (a) Calculated unitary time-evolution of the $^{13}$C nuclear spin in a vicinity of $B_P^{(0)}$; a similar response is obtained near $B_P^{(\pm 1)}$. In all cases, the system starts from a state where the NV$^-$ spin is in $m_S = 0$, its host $^{14}$N nuclear spin is in $m_J = 0$, and all other spins are in thermal equilibrium. (b) Modeled $^{13}$C NMR signal as a function of $B_P$ starting from a state where all spins (including the $^{14}$N at the NV) are in thermal equilibrium. This model also takes into account cross-relaxation with P1 centers both parallel and non-parallel with the field. The starker solid line is a convolution of the calculated line pattern with a 0.1 mT Gaussian. The P1–$^{13}$C coupling tensor is listed in the text; the NV$^-$– P1 coupling is 500 kHz.

Here, the original hyperfine term in Eq. (1) that coupled the NV to the $^{13}$C has been replaced by $\boldsymbol{S'} \cdot \boldsymbol{A}_{C'} \cdot \boldsymbol{I'}$, with $\boldsymbol{A}_{C'}$ denoting the P1-$^{13}$C hyperfine tensor, and $\boldsymbol{I'}$ is the carbon spin operator. In particular, we use $A_{C',zz} = 14$ MHz, $A_{C',xx} = A_{C',yy} = 10$ MHz, and an anisotropic term $A_{C',zx} = A_{C',xz} = 4$ MHz. These values are consistent with those reported in the literature.[23,24]

Supp. Fig. 1a shows the calculated unitary evolution of the $^{13}$C nuclear spin for a particular range of the magnetic field $B_P$. When comparing to the dynamics shown in Fig. 4(b) of the main text, find that the polarization transfer is now ~4 times slower, though substantial levels of polarization (exceeding 4%) can be attained even without further optical pumping. Supp. Fig.

1-b shows the projected amplitude of the $^{13}$C spin polarization as a function of the field $B_P$. While the global structure of the lines is similar to that obtained with the NV-$^{13}$C interaction, a notable difference is the narrow separation between the positive and negative peaks (polarizing and anti-polarizing, respectively). This separation, which in the case of the NV-$^{13}$C interaction roughly corresponds to $|A_{C,zz}|/2 \sim 7$ MHz, is now much smaller (~500 kHz). In fact, this difference arises from the spin-1 nature of the NV$^-$, whose $m_S = \pm 1$ states are split by $|A_{C,zz}|$ whereas the $m_S = 0$ state remains (to first order) immune to the hyperfine coupling to the $^{13}$C. By contrast, both projections of the P1 — a spin-1/2 system — are *symmetrically* split by the hyperfine coupling to the $^{13}$C.

The stark solid line in Suppl. Fig. 1c shows the (re-scaled) convolution of the projected polarization amplitudes with a Gaussian of 2.7 MHz (same as in the main text). Despite a visible assymetry between the positive and negative extrema, the pattern that emerges still exhibits a resemblance with our experimental observations. Given the larger concentration of P1 centers — ten or more times more abundant than the NVs — it is presently difficult to rule out P1-coupled carbons as contributing to the measured polarization.